\documentclass{vgtc}                          % final (conference style)
\usepackage{mathptmx}
\usepackage{times}
\usepackage{epsfig}
\usepackage{graphicx}
\usepackage{amsmath}
\usepackage{amssymb}
\usepackage[linesnumbered,ruled,vlined]{algorithm2e}
\usepackage{booktabs}
\usepackage{array}

\onlineid{0}

\vgtccategory{Research}

\vgtcinsertpkg

\title{3D visual analysis of seabed on smartphone}

\author{Zhihan Lv$^{1}$, Tianyun Su$^{2}$, Xiaoming Li$^{1,3,4}$, Shengzhong Feng$^{1}$\\
1. Shenzhen Institutes of Advanced Technology(SIAT), Chinese Academy of Science, China\\
2. The First Institute of Oceanography(FIO), State Oceanic Administration(SOA), China\\
3. Shenzhen Research Center of Digital City Engineering, Shenzhen, China\\
4. Key Laboratory of Urban Land Resources Monitoring and Simulation, Ministry of Land and Resources, Shenzhen, China\\
lvzhihan@gmail.com, sutiany@fio.org.cn, sz.feng@siat.ac.cn} %Tonglin属于第2个单位，同时还是通讯作者

\abstract{
We create a 'virtual-seabed' platform to realize the 3D visual analysis of seabed on smartphone. The 3D seabed platform is based on a 'section-drilling' model, implementing visualization and analysis of the integrated data of seabed on the 3D browser on smartphone. Some 3D visual analysis functions are developed. This work presents a thorough and interesting way of presenting seabed data on smartphone,
   which raises many application possibilities. This platform is another practical proof based on our WebVRGIS platform.
} % end of abstract

\CCScatlist{ 
 \CCScat{1.3.7}{Computer Graphics}{Three-Dimensional Graphics and Realism}{Virtual reality}
}

\begin{document}

%% The ``\maketitle'' command must be the first command after the
%% ``\begin{document}'' command. It prepares and prints the title block.

%% the only exception to this rule is the \firstsection command

\maketitle

\begin{figure}
    \begin{center}
    \includegraphics[width=1\columnwidth]{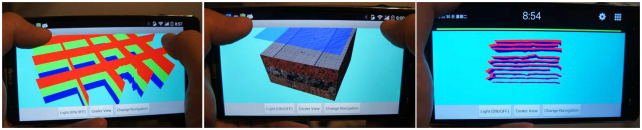}
    \caption{3D visual analysis of seabed on smartphone }
    \label{fig:seabed7}
        \end{center}
\end{figure}

\section{Introduction}

'Digital Seabed', as a hotspot of the marine information technology, the best way to represent its information to participants is to convey and perform marine scenes in an intuitive way~\cite{Chen20102524}. The features of 3D visualization include intuitiveness, time sharing, and regularity. Besides, it is good for displaying internal laws and different relationships in the data. It is a challenge to integrate the visualization of massive marine data into the virtual environment organically. Beside spatial data integration, new user interfaces for seabed geo-database is also expected~\cite{Breunig:2011:RGR:1998664.1998871}. The proposed system has reserved the real-time features of interactive roaming and analysis on mobile device, as shown in Figure~\ref{fig:seabed7}. The applications of virtual and mixed-reality environments, in fields like simulation, games and education, bring in billions of dollars every year~\cite{5582532}. The utilization of these virtual and mixed-reality environments is under exploration for various tools for research or commerce.
Besides, with the development of marine information technology, multi-dimensional dynamic visualization of the water environment data becomes a hotpot in marine research, which is mainly reflected in simulation and emulation of the information on water environmental elements including sea temperature, ocean current salinity and seawater density. An exploratory research about 3D ocean current model rendering and multi-touch interaction~\cite{6107079} was conducted early. The visualized analysis based on the time-space characteristic of marine environment data has been already implemented on web context~\cite{5567751}. The integrated virtual reality system of marine environment has been utilized for geospatial analysis on high performance computer~\cite{Li:2011:VSI:2048601.2048634}. However, all of current related systems are dependent on the high performance computer, since they are using complexity algorithms to realize a realistic virtual-seabed.

\begin{figure}
    \begin{center}
    \includegraphics[width=1\columnwidth]{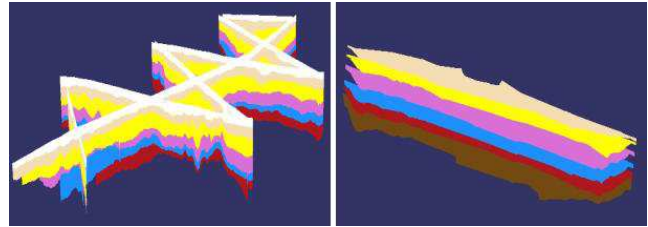}
    \caption{Left: Grid-like figure; Right: Layer diagram}
    \label{fig:seabed3}
        \end{center}
\end{figure}

\section{Approach and Results}

Our aim, instead of just building a virtual environment replicating the seabed, is to create a ubiquitous platform which can collect data of the variance of strata beneath the seabed, and then find efficient approaches (i.e. analysis and visualization) to use and display that information for various phenomena. The core data structure is so-called 'section-drilling' model, which is based on the features of the comprehensive data from multiple sources under the seabed. The complete description of 'section-drilling' model is introduced at~\cite{su20143d}. Based on the core algorithm, a component is extended to smartphone ubiquitous context, the mobile prototype has been presented at Siggraph Asia 2014~\cite{lv20143d}. Our cooperative network architecture supports the data flow~\cite{lviconip}, which is based on WebVR engine~\cite{lv2011webvr}. 

On smartphone browser, an application based on the 'section-drilling' model demonstrate its omnipresent feature. The consequent '3D Seabed' is applied to fields like simulation, visualization, and analysis; a set of interlinked, real-time layers that capture the information about the 3D Seabed, as well as its analysis result, help to accomplish that. Comparing to our previous work, this paper extends the seabed terrain to water body and water temperature visualization.

\begin{figure}
    \begin{center}
    \includegraphics[width=1\columnwidth]{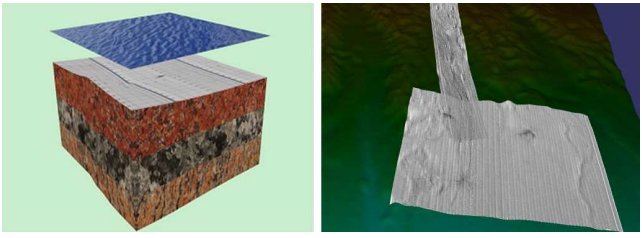}
    \caption{Left: 3D seabed; Right: Sidescan sonar image and multibeam fusion renderings}
    \label{fig:seabed4}
        \end{center}
\end{figure}

\begin{figure}
    \begin{center}
    \includegraphics[width=1\columnwidth]{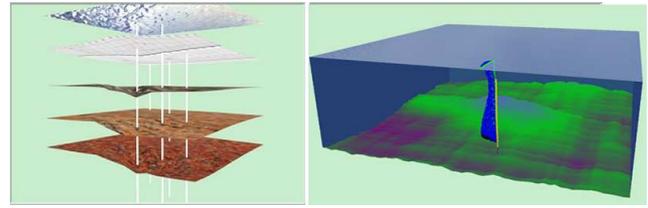}
    \caption{Left: Drilling correction visualization; Right: Oil spilling dynamic visualization}
    \label{fig:seabed5}
        \end{center}
\end{figure}

The grid-like figure in Figure~\ref{fig:seabed3} is built by delaunay triangulation algorithm. The converted coordinate sequence has hierarchical information. The fence diagram is obtained by hierarchical rendering of stratums according to the survey line. The horizon diagram is obtained by triangular processing of discrete peaks in the same stratum. The sidescan sonar image and multibeam fusion renderings function employed the undersea canyon area data to present, as shown in Figure~\ref{fig:seabed4}. The oil spilling dynamic visualization in Figure~\ref{fig:seabed5} right is implemented by 3D dynamic particle system.

\begin{figure}
    \begin{center}
    \includegraphics[width=1\columnwidth]{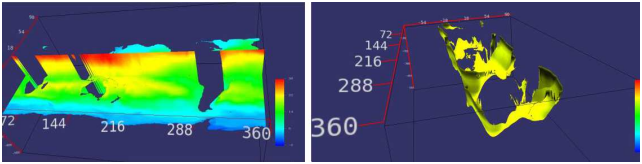}
    \caption{Left: Sectional temperature map 200 meters under water in 0.5'N. latitude; Right: Marine water environmental temperature isosurface on 20 degrees}
    \label{fig:seabed2}
        \end{center}
\end{figure}

In addition, some visual analysis methods have been implemented and worked on desktop computer smoothly, but haven't been ported to mobile device yet. Nonetheless, all the algorithms are designed taking into account computational efficiency on ubiquitous context. As shown in Figure~\ref{fig:seabed2} left, the sectional temperature map and marine water temperature field internal profile are visualized by volume rendering based on 3D texture. The transparency of invalid data point is set as 0 by adding judgment identification. Moreover, the internal section of parameter rendering volume is rendered through setting volume. The marine water environmental temperature Isosurface in Figure~\ref{fig:seabed2} right is rendered by improved Marching Cubes algorithm. To improve the drawing quality, the normal vector of triangular mesh on the contour surface is obtained based on the estimation method of the gradient normal vector, so as to realize the smoothness of the contour surface. 

The client is wrapped into a component extending to smartphone ubiquitous context. The component is implemented in C++, Java with Android SDK/NDK, the 3D rendering tasks were realized using OpenGL ES.

\section{Conclusions}
Agreed as being both directly useful and generally extendable for future applications, the mature criterion application is a beneficial tool for the scientists and public for browsing and analyzing the information from the seabed intuitively. Three-dimensional or multi-dimensional visualization of spatial information is a popular and tough topic. The spatial model and criterion application, the topic of this paper, are still under discussion and remain a primary research. Many key problems, for example, the visualization of dynamic data reflecting the particular geological phenomenon indicating faults and folds, still need to be researched and worked out in the future. We believe there remains more value for a broader field, especially in a utility seabed monitoring device that provides unremitting multi-source and multi-dimensional synthesis of virtual and ubiquitous applications. In our future work, in order to improve the system performance, abstract visualization approach~\cite{zhong2012spatiotemporal} is considered to represent the ocean current data. To enhance the flexibility of user-and-system interaction, a novel touch-less technology is considered as the input method~\cite{lv2013wearable}. Beside scientific field, we plan to perform a virtual seabed community project based on this framework~\cite{lv2012framework} in civil field, as a rich mobile application~\cite{zhang2009research}, to provide a intuitive and usable communication tool for ocean field experts.

\section{Acknowledgments}
The authors are thankful to the National Natural Science Fund for the Youth of China (41301439) and Shenzhen Scientific \& Research Development Fund (CXZZ20130321092415392).

\bibliographystyle{abbrv}
%%use following if all content of bibtex file should be shown
%\nocite{*}
\bibliography{VRVIDEO}
\end{document}